# Effects of excess carriers on native defects in wide bandgap semiconductors: illumination as a method to enhance p-type doping

Kirstin Alberi[1,*] and Michael A. Scarpulla[2,3]
[1] National Renewable Energy Laboratory, Golden, CO, USA
[2] Materials Science & Engineering, University of Utah, Salt Lake City, UT, USA
[3] Electrical & Computer Engineering, University of Utah, Salt Lake City, UT, USA

**Abstract**

Undesired unintentional doping and doping limits in semiconductors are typically caused by compensating defects with low formation energies. Since the formation energy of a charged defect depends linearly on the Fermi level, doping limits can be especially pronounced in wide bandgap semiconductors where the Fermi level can vary substantially. Introduction of non-equilibrium carrier concentrations during growth or processing alters the chemical potentials of band carriers and thus provides the possibility of modifying populations of charged defects in ways impossible at thermal equilibrium. Herein we demonstrate that, for an ergodic system with excess carriers, the rates of carrier capture and emission involving a defect charge transition level rigorously determine the admixture of electron and hole quasi-Fermi levels determining the formation energy of non-zero charge states of that defect type. To catalog the range of possible responses to excess carriers, we investigate the behavior of a single donor-like defect as functions of extrinsic doping and energy of the charge transition level. The technologically most important finding is that excess carriers will increase the formation energy of compensating defects for most values of the charge transition level in the bandgap. Thus, it may be possible to overcome limitations on doping imposed by native defects. Cases also exist in wide bandgap semiconductors in which the concentration of defects with the same charge polarity as the majority dopant is either left unchanged or actually increases. The causes of these various behaviors are rationalized in terms of the capture and emission rates and guidelines for carrying out experimental tests of this model are given.

* Emails: Kirstin.Alberi@nrel.gov, Mike.Scarpulla@utah.edu

**Introduction**

Controlling native defect populations in semiconductors remains a longstanding and critical endeavor in our effort to advance optoelectronic device performance. Extrinsically-doped wide bandgap semiconductors, in particular can suffer acutely from high native defect concentrations that inherently form according to the "doping-limit rule".[1,2] It is well-known that the formation energy of a charged point defect depends on the Fermi level due to the real exchange of charged carriers with either the valence or conduction band. Shifting the Fermi level toward one band edge through extrinsic doping causes the formation energies of all possible compensating defects in the system to decrease. While the concentration of extrinsic dopants is controlled, the number of sites available to form native compensating defects is in the vast majority of cases at least 100 times higher. Additionally, for many extrinsic dopants, self-compensation may occur because there are multiple sites and configurations that can compensate the typical substitutional dopant configuration. These effectively infinite reservoirs of compensating defects lead to limits on the net free carrier concentrations achievable. The particular limits are specific to the set of native and self-compensating extrinsic defects in the semiconductor.

The larger energy range over which the Fermi level can vary in wide bandgap semiconductors generally results in lower doping limits compared to semiconductors with smaller bandgaps. In some cases, including ZnSe[3,4], GaN[5,6] ZnO[7] and many other oxide semiconductors with extrinsic dopants added in attempts to achieve p-type doping the compensating native defects can overwhelm intentionally added shallow dopants, resulting in very low or even negligible extrinsic free carrier concentrations. The fundamental thermodynamic link between the Fermi level and the equilibrium defect formation energy means that crystal growth or processing conditions at thermal equilibrium cannot overcome these doping limits. Thus some aspect of thermal and mass equilibrium (i.e. the reversible exchange of thermal energy or mass between the semiconductor and external reservoirs) must be broken in order to overcome these doping limits.

Semiconductor processing involving mass exchange between a vapor or liquid phase and the solid crystal (e.g. annealing, molecular beam epitaxy, organometallic vapor phase epitaxy, bulk crystal growth, liquid phase epitaxy, etc.) are controlled by N+1 thermodynamic potentials: the chemical potentials for the N species exchanged and temperature. The chemical potentials embody the partial pressures and excited internal degrees of freedom of arriving species. For charged defects, the Fermi level is self-consistently determined with the intrinsic and extrinsic defect concentrations by enforcing charge balance. Splitting the quasi-Fermi levels (QFL), which implies the flow of energy other than heat into or out of the semiconductor region in question (e.g. via absorption of light, x-ray or other forms of ionizing radiation or by current injection) brings the total number of thermodynamic potentials and conjugate concentrations to N+2. Manipulation of the QFLs can thus be expected to offer a way to extend the doping limits in at least some semiconductors. Due to the large possible dynamic range of the QFLs, the effects in wide bandgap semiconductors can potentially lead to especially dramatic changes.

We have recently demonstrated theoretically that non-equilibrium carrier concentrations, by definition those in excess or deficit of the thermal equilibrium concentration, represent an additional thermodynamic potential or control variable in semiconductor processing.[8] Our theoretical framework self-consistently determines steady-state QFL splitting and concentrations of defects by simultaneously considering charge balance and detailed balance of carriers between the bands and all possible charge transition levels. In this prior work, we utilized n-type GaSb as a model system for simplicity because its defect equilibrium is dominated by one compensating native defect with low formation energy (the Ga antisite $Ga_{Sb}$). We demonstrated that the $Ga_{Sb}$ concentration can be suppressed by several orders of magnitude even for low-injection conditions. However, the rather narrow bandgap of GaSb (0.7 eV) limits the dynamic range of the effect. We also restricted the investigation to the specific calculated formation energies, charges, and charge transition energies calculated for the common defects in GaSb – i.e. we did not fully investigate the general behavior of this model for coupling of excess carriers to defect formation energies. Considering the much greater QFL splitting possible in wide bandgap semiconductors, harnessing excess carriers to reduce compensating defects and raise doping limits may yield much more dramatic and useful results. Such an approach would enable higher performance devices and even true homojunctions in cases where significant doping of one polarity is ephemeral and elusive (e.g. p-type ZnO). It is therefore important to understand the types of behaviors possible for different classes of defects.

In this work we investigate the behavior of both majority and compensating trap states as functions of charge transition levels, cross-sections, excess carrier concentrations, and temperature for wide bandgap semiconductors in order to determine the classes of behavior expected for various general cases. We have considered only cases in which the chemical potentials of elements relevant to the defects are specified; we have not included results for cases where the concentration of elements forming the defects considered are constrained. For a generic defect with 0 and +1 charge states, we find that the factor having the strongest influence on how the defect formation energy changes with QFL splitting is the position of the charge transition level, $E_t$, in the bandgap. The contributions of each QFL to the defect formation energy depend on the relative weighting of the rates of carrier exchange with the bands and these in turn are sensitive to $E_t$. When the carrier emission rate involved in the defect formation process is high, compensating defect concentrations can be reduced. Yet, when carrier capture (which is not thermally activated) dominates, the concentration of defects having the same charge as the majority extrinsic dopant may actually *increase*. The general behaviors detailed herein establish guidelines for understanding how excess carriers would affect the concentrations of various specific defects in specific materials during growth or processing.

**Theoretical Approach**

We begin by detailing the general theoretical framework, which we have elaborated slightly since our last publication.[8] We consider a binary compound semiconductor AB uniformly doped with a fixed density of shallow extrinsic acceptors $N_A$. We consider only native defects herein; i.e. we do not consider the case where the extrinsic dopants may form defects other than the shallow acceptors. The equilibrium concentration of a native defect $X$ in the charge state $q$ can be expressed as[9]:

$$[X^q] = g N_{sites} \exp(-\Delta H_f^q / k_B T) \tag{1}$$

in which $g$ is the number of equivalent configurations, $N_{sites}$ is the density of sites on which the defect can be located, $T$ is the temperature, $k_B$ is the Boltzmann constant, and $\Delta H_f^q$ is the formation energy for the charged defect. $\Delta H_f^q$ is subsequently expressed as:

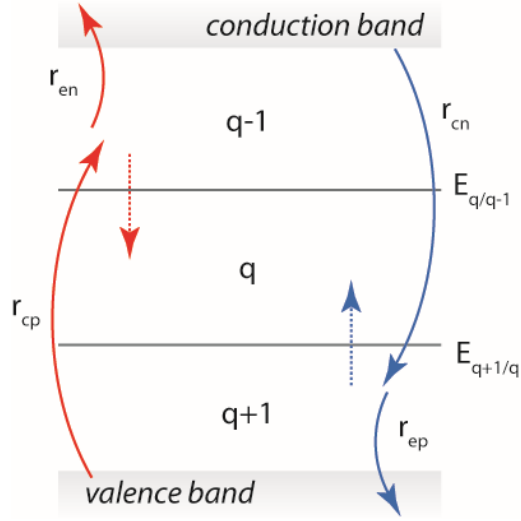

**Fig. 1** (Color online) Schematic of the electron and hole ( n or p) emission and capture (e or c) rates involved in creating a defect with charge q from defects with charges q+1 or q-1.

$$\Delta H_f^q = E_D + q(\mu_F + E_{VB}) - n_A\mu_A - n_B\mu_B \tag{2}$$

in which $E_{VB}$ is the energy of the valence band maximum and $E_D$ is the total energy of the supercell comprised of $n_A$ and $n_B$ atoms and a single charged defect. Charge $q$ is added to or subtracted from the defect by carrier exchange with the conduction or valence band edges. The Fermi level, which we write as $\mu_F$ to highlight its role as the chemical potential for carriers, is therefore included to characterize the chemical potentials of these free carrier reservoirs. The electron QFL $\mu_{Fn}$ gives the chemical potential for electrons in the conduction band and the hole QFL $\mu_{Fp}$ gives the chemical potential for holes in the valence band. In thermal equilibrium with no excess carriers, the bands are in equilibrium and thus $\mu_{Fn} = \mu_{Fp} = \mu_F$. Finally, $\mu_A$ and $\mu_B$ are the chemical potentials of atoms *A* and *B*. Their values are limited to be less than those of their bulk phases, $\mu_{A(bulk)}$ and $\mu_{B(bulk)}$ in order to ensure that only the single AB phase exists. The corresponding chemical potential of the AB compound is assumed to be the sum of $\mu_A$ and $\mu_B$ such that Eq. 2 can be re-written as:

$$\Delta H_f^q = E_D' + q\mu_F - \frac{1}{2}(n_A - n_B)\Delta\mu \tag{3}$$

in which

$$E_D' = E_D - \frac{1}{2}(n_A + n_B)\mu_{AB} - \frac{1}{2}(n_A - n_B)(\mu_{A(bulk)} - \mu_{B(bulk)}) + qE_{VB} \tag{4}$$

and

$$\Delta\mu = (\mu_A - \mu_B) - (\mu_{A(bulk)} - \mu_{B(bulk)}) \tag{5}$$

In the presence of non-equilibrium photogenerated carrier populations, the electron and hole chemical potentials are characterized separately by QFLs $\mu_{Fn}$ and $\mu_{Fp}$ instead of $\mu_F$. Their relative contributions to the defect formation energy of a defect in a particular charge state should be determined by the possible carrier capture and emission processes with the valence and conduction bands. For example, electron capture from the q+1 or electron emission from the q-1 charge states both involve µFn as the relevant carrier chemical potential and will both result in the q charge state, as shown in Fig. 1. For a donor-like defect, only the q=0 and +1 states exist and thus only electron emission or hole capture are possible for the q=0 state while hole emission or electron capture are possible for q=+1.

The detailed balance rate equations describing the capture processes are[10,11,12]:

$$r_{c,n} = v_n \sigma_n N_C F_{1/2}\left(\frac{\mu_{Fn} - E_C}{k_B T}\right) N_t^{q+1}$$

$$r_{c,p} = v_p \sigma_p N_V F_{1/2}\left(\frac{E_V - \mu_{Fp}}{k_B T}\right) N_t^{q-1} \tag{6a}$$

Likewise, the emission rates are:

$$r_{e,n} = v_n \sigma_n N_C \exp\left(\frac{E_t - E_C}{k_B T}\right) N_t^{q-1}$$

(6b)

$$r_{e,p} = v_p \sigma_p N_V \exp\left(\frac{E_V - E_t}{k_B T}\right) N_t^{q+1}$$

The parameters $v_n$ and $v_p$ are the band carrier thermal velocities, $\sigma_n$ and $\sigma_p$ are the capture cross sections, $N_C$ and $N_V$ are the effective densities of states, $\mathbb{F}_{1/2}(\eta)$ is the Fermi-Dirac function, and $E_C$ and $E_V$ are the energies of the conduction and valance band edges. $N_t^{q-1}$ and $N_t^{q+1}$ are the concentrations of defects starting in the charge states $q+1$ and $q-1$.

The ratios of the capture and emission rates for holes involving $\mu_{Fp}$ and for electrons involving $\mu_{Fn}$ describe the relative influence of each QFL on the charge formation process and will therefore determine their weighting. While only one process may occur at a time to move a specific, isolated defect into the q charge state, the relative magnitudes of the possible transition rates will determine the fractions of defects undergoing each possible transition in an ergodic system of defects. Thus, generalizing the formulation in our prior publication, we write the expectation value of $\mu_F$ for an ensemble as:

$$\mu_F = \left(\frac{r_{c,n} + r_{e,n}}{r_{c,n} + r_{e,n} + r_{c,p} + r_{e,p}}\right)\mu_{Fn} + \left(\frac{r_{c,p} + r_{e,p}}{r_{c,n} + r_{e,n} + r_{c,p} + r_{e,p}}\right)\mu_{Fn} \qquad (7)$$

This formula reproduces the required limiting behaviors. First, $\mu_F$ reverts to $E_F$ in the case of thermal equilibrium. For $k_B T \ll E_g$, $\mu_F \approx \mu_{Fn}$ for shallow donors and $\mu_F \approx \mu_{Fp}$ for shallow acceptors. For a charge transition level coincident with the intrinsic Fermi level $E_t = E_i$ and having equal products of cross section, thermal velocity, and effective density of states for electrons and holes ($v_n \sigma_n N_C = v_p \sigma_p N_V$), the electron and hole QFLs contribute equally to $\mu_F$.

Note that this very specific special case corresponds to a previous framework presented in Ref.13. Because of the exponential dependencies of the emission rates on $E_t$, the admixture of QFLs is determined in large part by whether $E_t$ is closer to $E_C$ or $E_V$.

We now consider a situation of a semiconductor crystal in thermal equilibrium with reservoirs of different atoms but exposed to a steady-state flux of energy (e.g. irradiation with above bandgap photons) that produces an electron and hole generation rate G. The defect formation energies for each charge state are typically computed from first principles calculations, allowing a set of N equations for the N possible $X_i^q$ in the system to be expressed only in terms of the two QFLs. The set of N equations of form Eq. 3 for the defect populations are determined self-consistently with two additional equations: the charge balance (Eq. 8) and steady-state excess carrier population conditions (Eq. 9). The charge balance condition is:

$$\sum_i q[X_i^q] + N_D^+ - N_A^- - n + p = 0 \tag{8}$$

In steady-state, the generation ($G$) and net recombination ($U$, which is the recombination only of excess carriers) rates balance such that:

$$\begin{aligned}\frac{dn}{dt} &= G - U_{BB} - U_{Aug} - \sum_i U_{SRH}\left([X_i^q]\right) = 0 \\ \frac{dp}{dt} &= G - U_{BB} - U_{Aug} - \sum_i U_{SRH}\left([X_i^q]\right) = 0\end{aligned} \tag{9}$$

In Eqs. 8 and 9, $i$ indexes over all defect types, $p = p_o + \Delta p$ and $n = n_o + \Delta n$ are the total carrier densities including the excess holes $\Delta p$ and electrons $\Delta n$, $U_{BB}$ is the band-to-band recombination rate, $U_{Aug}$ is the Auger recombination rate, and the $U_{SRH}$ terms account for Shockley-Read-Hall recombination involving each of the defects' charge transition levels. The net recombination rates in Eq. 9 are expressed as:

$$U_{BB} = \beta_{BB}\left(np - n_i^2\right)$$

$$U_{Aug} = \left(C_{Aug,n} n + C_{Aug,p} p\right)\left(np - n_i^2\right)$$

$$U_{SRH}\left(\left[X_i^q\right]\right) = \frac{np - n_i^2}{\frac{1}{\sigma_p v_p \left[X_i^q\right]}\left(n + N_C \exp\left[\frac{-(E_{CB} - E_t)}{k_B T}\right]\right) + \frac{1}{\sigma_n v_n \left[X_i^q\right]}\left(p + N_V \exp\left[\frac{-(E_t - E_{VB})}{k_B T}\right]\right)}$$

(10)

in which $\beta_{BB}$ is the band-to-band recombination coefficient, and $C_{Aug,n}$ and $C_{Aug,p}$ are the electron and hole Auger coefficients. All of the quantities in Eqs. 11 and 12 are expressed in terms of $\mu_{Fn}$ and $\mu_{Fp}$, allowing a self-consistent solution for the N+2 unknowns to be found for each set of conditions. In our solution procedure we take into account the temperature dependencies of all quantities whenever possible. We have to date utilized the Boltzmann approximation for band carrier densities and thus have omitted degenerate cases from final results.

**Generic Defect Calculations**

Examination of equations 1-7 shows that the Fermi level plays a dominant role in determining the formation energy and concentration of defects with non-zero charge. When the QFLs are split by the introduction of non-equilibrium carrier concentrations, the contribution of $\mu_{Fn}$ and $\mu_{Fp}$ are governed by the relative magnitudes of the electron and hole capture and emission rates that contribute to the generation of a specific charge state. Consider the case of forming a generic donor-like defect with +1 charge, $X^{+1}$, from its charge-neutral state, $X^0$. This process occurs when $X^0$ either emits an electron or captures a hole (see Fig. 1). Only these two rates ($r_{e,n}$ and $r_{c,p}$, respectively) are relevant for determining the admixed $\mu_F$ controlling the formation enthalpy; $r_{e,p}$ and $r_{c,n}$ play no role.

The electron emission rate $r_{e,n}$ predominantly depends on the energetic separation of the charge transition level, $E_t = E_{+1/0}$, from the conduction band edge. The hole capture rate predominantly depends on the hole concentration p and thus on the extrinsic doping concentration, temperature-dependent intrinsic carrier density $n_i$, and photogeneration rate. Whether $r_{e,n}$ or $r_{c,p}$ is the dominant rate therefore depends on the specific $E_{+1/0}$ and free carrier concentration values. Furthermore, if $\mu_F$ increases, the formation enthalpy of $X^{+1}$ will increase and thus its concentration $[X^{+1}]$ will decrease.

If $r_{e,n} >> r_{c,p}$ such that $\mu_F$ largely follows $\mu_{Fn}$ and if $\mu_{Fn}$ increases under photogeneration (e.g. for low p-type doping and $E_{+1/0}$ near the conduction band ), then $[X^{+1}]$ will decrease. On the other hand, if $r_{c,p} >> r_{e,n}$ such that $\mu_F$ largely follows $\mu_{Fp}$ and if $\mu_{Fp}$ decreases under photogeneration (e.g. for low n-type doping and $E_{+1/0}$ far away from the conduction band), then $[X^{+1}]$ will actually *increase*. Finally, if there is no change in the QFL that controls $\mu_F$ (for example if temperature is sufficiently high that the excess carrier lifetime is small and intrinsic carriers dominate or under heavy extrinsic doping), then no change in $[X^{+1}]$ is expected for G>0 compared to the thermal equilibrium ($G$=0) conditions.

To understand the ranges of behaviors and distinct physical cases embodied in the model for defects having only two charge states, we simulated a hypothetical donor-like native defect in a semiconductor having bandstructure properties mimicking those of ZnSe.[14] Of course, the qualitative cases for an acceptor-like defect can be obtained if the extrinsic doping polarity is also inverted. Calculations were performed over a range of extrinsic doping conditions from $10^{17}$ cm$^{-3}$ n-type to $10^{17}$ cm$^{-3}$ p-type and temperatures from 200 – 600 °C intended to span a wider range of conditions than those typically applied during molecular beam epitaxy (MBE) growth. The ZnSe bandgap temperature dependence was obtained from Ref. 15.

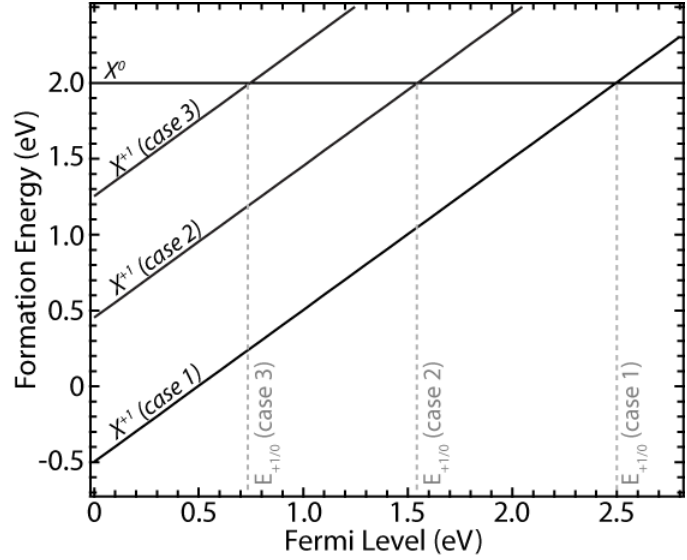

**Fig. 2** Formation energies of a hypothetical donor-like defect having +1 and 0 charge states as functions of dark Fermi level ranging across the ZnSe bandgap at 200 °C. The charge transition levels $E_t=E_{+1/0}$ for cases 1-3 are the Fermi levels at which the neutral and +1 formation enthalpies are equal.

Since the position of $E_{+1/0}$ relative to the conduction band substantially affected $r_{e,n}$ and thus the value of the admixed $\mu_F$, we explored three cases in which $E_{+1/0}$ was (1) close to $E_C$, (2) in the middle of the bandgap and (3) close to $E_V$. These cases are represented in Fig. 2 for a substrate temperature of 200 °C, the lowest temperature considered. We chose to keep this energy separation constant as a function of substrate temperature so that it would be easier to observe its contribution to the defect behavior. This was achieved by assuming that 2/3 of the bandgap reduction with increasing temperature occurred because of downwards shifting of $E_C$ and 1/3 occurred because of $E_V$ increasing. We then allowed $E_{+1/0}$ to follow $E_C$ for all temperatures (i.e. we kept $E_c$-$E_{+1/0}$ constant versus temperature for this donor-like defect). We arbitrarily set the formation enthalpy of the neutral charge state to 2.0 eV. The electron and hole capture cross sections of the $X^0$ state were both set to $\sigma_n=\sigma_p=10^{-16}$ cm$^2$, while those of the $X^{+1}$ state were set to $\sigma_n=10^{-14}$ and $\sigma_p=10^{-17}$ cm$^2$ (which would be characteristic of a pure Coulomb potential). In addition to with "dark" thermal equilibrium conditions, we studied three photocarrier generation rates G=$10^{13}$, $10^{16}$ and $10^{19}$ cm$^{-3}$s$^{-1}$. We ensured that the system remained non-degenerate for all temperatures (this constrained G<$10^{20}$ cm$^{-3}$s$^{-1}$ at 600 °C). For the worst case scenario, assuming a maximum carrier lifetime on the order of 1000 ns for intrinsic material, these generation rates would produce excess carrier concentrations between $10^7$ and $10^{13}$ cm$^{-3}$, i.e. in low-injection for the extrinsic dopings considered. However, since our formulation does not invoke the lifetime approximation, cases of high injection may also be studied provided that degeneracy of the quasi-Fermi levels is not reached or full Fermi-Dirac statistics are used for the bands.

The calculated values of the carrier concentrations, $r_{e,n}$ and $r_{c,p}$, $\mu_{Fn}$ and $\mu_{Fp}$ and [$X^{+1}$] as a function of the growth temperature are displayed in Fig. 3 for Case 1, where $E_{+1/0}$ is close to $E_C$. Only three of the seven doping concentrations examined in this study are shown in the figure for improved readability. Additional doping concentrations, as well as explicit calculations of the corresponding electron and hole free carrier concentrations are shown in Fig. S1 of the Supplementary Materials section. The [$X^{+1}$] defect concentration, shown in Fig. 3a, is determined largely by $r_{e,n}$ over $r_{c,p}$ under almost all doping and photogeneration conditions studied. [$X^0$], not shown, does not change with non-equilibrium carrier generation because its formation enthalpy does not depend on the Fermi level. However, it increases with temperature according to Eq. 1. For n-type doping ($10^{16}$ cm$^{-3}$) and low substrate temperatures, $\mu_{Fn}$ remains unchanged by the excess electrons injected by the relatively low photocarrier generation rates considered here, while $\mu_{Fp}$ moves downward with increasing $G$ (see Fig. 3b). At high temperatures, $\mu_{Fn}$ and $\mu_{Fp}$ begin to converge toward their intrinsic values. $r_{c,p}$ is particularly low because the hole concentration is small. Both $r_{e,n}$ and $r_{c,p}$, shown in Fig. 3d, increase with substrate temperature, as they both follow [$X^0$] according to Eq. 6. Overall, $\Delta H_f^{+1}$ is completely determined by $\mu_{Fn}$ in this situation, and because it does not change with G for the values considered, there is no change in [$X^{+1}$] with $G$. The increase in [$X^{+1}$] with substrate temperature is again a consequence of the temperature dependence in Eq. 1.

Non-equilibrium carrier concentrations start to have an effect as the extrinsic doping is changed to p-type such that $X^{+1}$ is a compensating defect. $r_{e,n}$ still greatly exceeds $r_{c,p}$ for modest p-type doping ($10^{14} - 10^{16}$ cm$^{-3}$), but $\mu_{Fn}$ now increases as $G$ increases. This results in an increase in $\mu_F$ and a decrease in [$X^{+1}$]. Only for the highest extrinsic p-type doping and photogeneration conditions considered herein does $r_{c,p}$ approach $r_{e,n}$ and thus temper the decrease in [$X^{+1}$] induced by the addition of excess carriers (see Fig. S1 in the Supplementary Materials section for more details). We reiterate that this case corresponds to a typical compensating donor-like defect with energy level close to $E_C$ in p-type material. $\Delta H_f^{+1}$ is quite low under such conditions (see Fig. 2a), and [$X^{+1}$] is thus high under equilibrium in the dark, meaning that it largely-compensates the extrinsic doping imposed. Consequently, for thermal equilibrium $\mu_F$ resides near mid-gap for doping concentrations up to $10^{17}$ cm$^{-3}$. This sets up the situation wherein even very low photogeneration rates can substantially move $\mu_{Fp}$, especially at low temperatures.

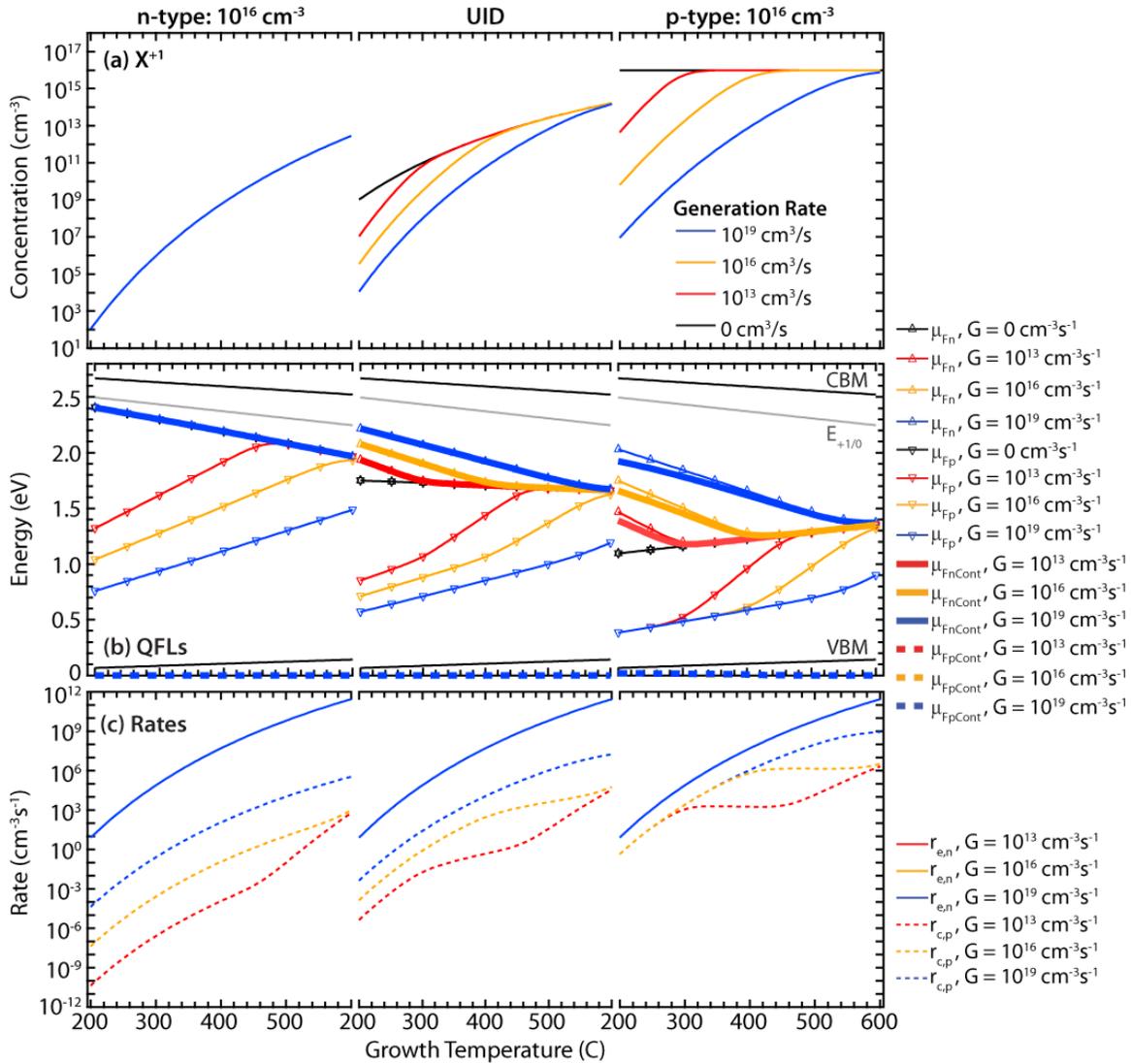

**Fig. 3** (Color online) Calculated parameters for Case 1 in which $E_{+1/0}$ is positioned close to $E_c$. Calculations were carried out as a function of temperature, extrinsic doping concentration (noted at the top of the figure) and photogeneration rate, *G*. UID stands for unintentionally doped. (a) Concentration of $X^{+1}$ defects, (b) $\mu_{Fn}$ and $\mu_{Fp}$ along with the contribution of $\mu_{Fn}$ (denoted $\mu_{FnCont}$) and $\mu_{Fp}$ (denoted $\mu_{FpCont}$) to $\mu_F$ used in the defect formation energy calculation. The CBM, VBM and $E_{+1/0}$ values are also marked. (c) Electron emission and hole capture rates. Full results including additional p and n-type doping concentrations are shown in Fig S1 of the Supplementary Materials section.

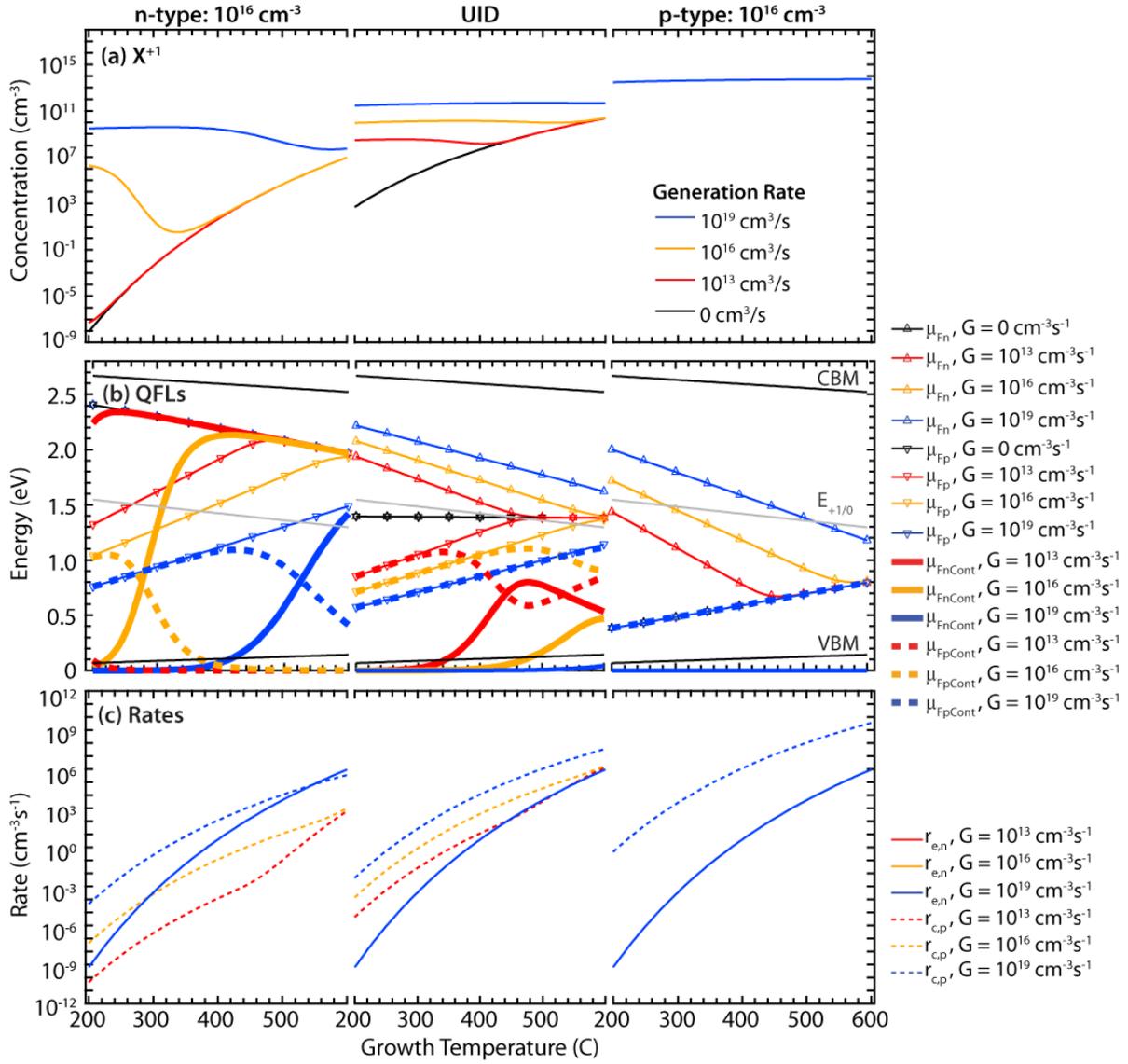

**Fig. 4** (Color online) Calculated parameters in the case where $E_{+1/0}$ is positioned near mid-bandgap. Calculations were carried out as a function of temperature, extrinsic doping concentration (noted at the top of the figure) and photogeneration rate, $G$. UID stands for unintentionally doped. (a) Concentration of $X^{+1}$ defects, (b) $\mu_{Fn}$ and $\mu_{Fp}$ along with the contribution of $\mu_{Fn}$ (denoted $\mu_{FnCont}$) and $\mu_{Fp}$ (denoted $\mu_{FpCont}$) to $\mu_F$ used in the defect formation energy calculation. The CBM, VBM and $E_{+1/0}$ values are also marked. (c) Electron emission and hole capture rates. Full results including additional p and n-type doping concentrations are shown in Fig. S2 of the Supplementary Materials section.

Figure 4 displays the same series of plots for Case 2 in which $E_{+1/0}$ is located near mid-gap. The much larger energetic difference between $E_{+1/0}$ and $E_C$ causes $r_{e,n}$ to be lower than $r_{c,p}$ under p-type conditions. The admixed $\mu_F$ under non-equilibrium conditions is thus determined by $\mu_{Fp}$, even for this donor-like defect (recall that "donor-like" refers only to the defect's set of possible charge states). $\Delta H_f^{+1}$ is also higher for Case 2 than for Case 1, such that $[X^{+1}]$ is no longer sufficient to fully compensate the imposed extrinsic p-type doping. $\mu_{Fp}$ and $[X^{+1}]$ therefore do not change for the modest photogeneration rates considered. As the doping becomes progressively more n-type, $\mu_{Fn}$ does start to contribute to the admixture of $\mu_{Fp}$. However, the decrease of $\mu_{Fp}$ with increasing $G$ leads to an overall decrease in $\mu_{Fp}$ and to an *increase* in $[X^{+1}]$ compared to equilibrium. To reiterate, for the charge transition level close to mid-gap and the case where the extrinsic doping polarity (i.e. donor-like) is the same as for the defect, its concentration can increase with increased excess carrier injection.

This behavior continues when $E_{+1/0}$ is positioned close to the $E_V$, shown for Case 3 in Fig. 5. The major difference is that $r_{c,p}$ and $\mu_{Fp}$ dominate at even higher n-type doping concentrations and lower $G$. Location of the $E_{+1/0}$ transition level near $E_V$ is possible for some deep defects, especially those with multiple charge transition levels. However, this is typically not the case for shallower donor states, which lie close to $E_C$.

We summarize here the most important findings from the above exploration of the regimes of behavior for the model assuming a single defect capable of being formed via exchange of atoms with external reservoirs and having only one charge transition level. When a native defect is compensating (i.e. a donor-like native defect in a semiconductor with extrinsic shallow acceptor dopants added), injection of non-equilibrium carriers will suppress the concentration of that defect within the mechanism of this model. This is particularly true in the case where $E_t$ is close to the band edge governing the relevant emission rate (i.e. $E_C$ for the case where electron emission controls the formation of the +1 charge state). Even when the energetic separation between the charge transition level and band is large, the change in the defect concentration with $G$ is expected to decrease rather than increase. Thus, injection of excess carriers e.g. by illumination during semiconductor growth or annealing is predicted by this model as a strategy for overcoming doping limits in such cases. The effects should be particularly large in wide-bandgap materials.

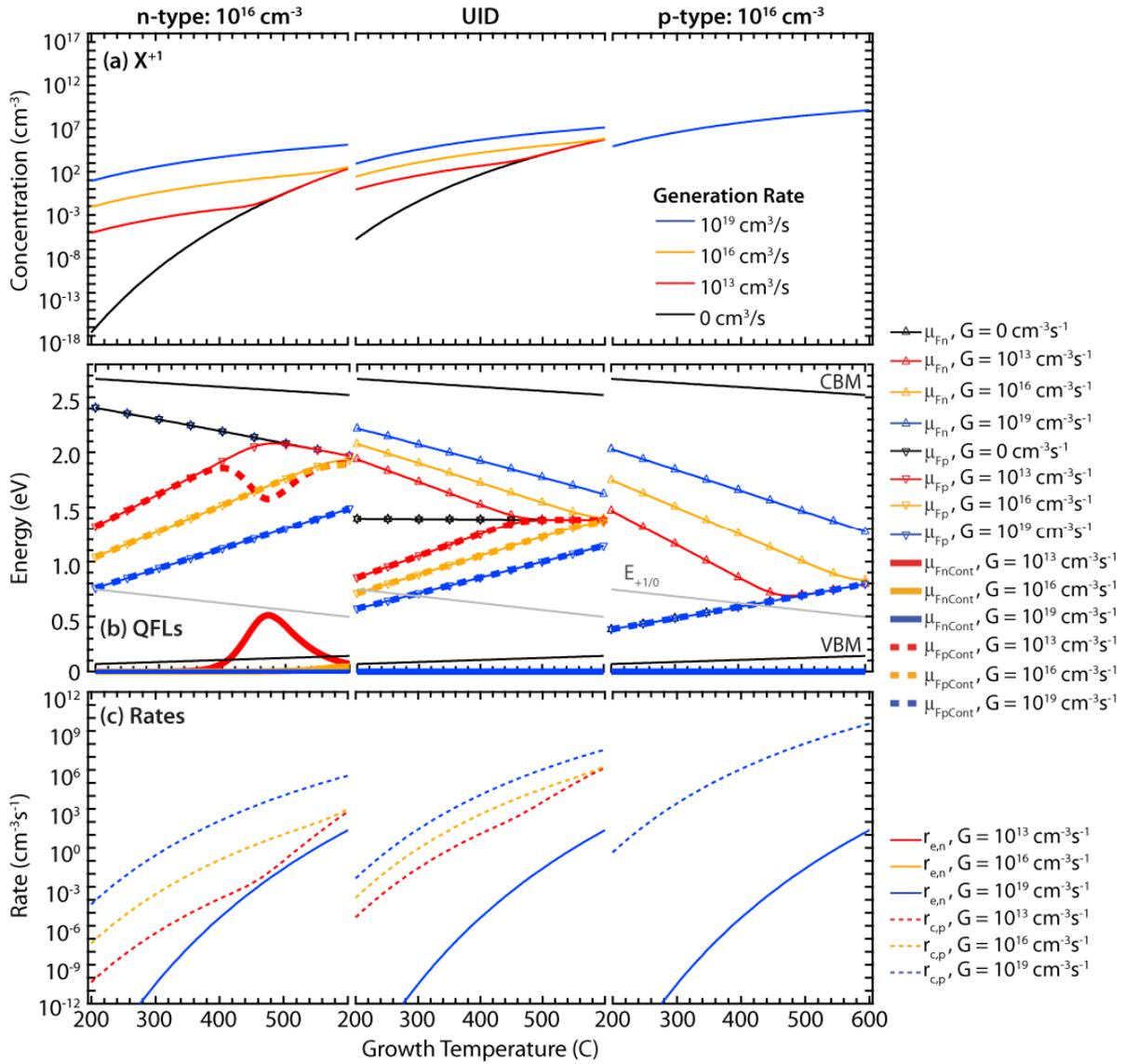

**Fig. 5** (Color online) Calculated parameters in Case 3 in which $E_{+1/0}$ is positioned near the VBM. Calculations were carried out as a function of temperature, extrinsic doping concentration (noted at the top of the figure) and photogeneration rate, $G$. UID stands for unintentionally doped. (a) Concentration of $X^{+1}$ defects, (b) $\mu_{Fn}$ and $\mu_{Fp}$ along with the contribution of $\mu_{Fn}$ (denoted $\mu_{FnCont}$) and $\mu_{Fp}$ (denoted $\mu_{FpCont}$) to $\mu_F$ used in the defect formation energy calculation. The CBM, VBM and $E_{+1/0}$ values are also marked. (c) Electron emission and hole capture rates. Full results including additional p and n-type doping concentrations are shown in Fig. S3 of the Supplementary Materials section.

When the defect is not compensating (i.e. a donor-like defect in an n-type semiconductor) and especially at low temperatures, excess carriers can cause an increase in the defect concentration. This behavior is enhanced in wide bandgap semiconductors because the QFL splitting can be large under reasonably small non-equilibrium carrier concentrations and the intrinsic carrier density is small. However, the defect concentration is typically much lower than in the case of compensating native defects, and the impacts of having an increased concentration of defects of the same polarity as the extrinsic dopant (at worst a majority carrier trap) may be easily outweighed by the ability to suppress a different, compensating defect in the system.

Specific dark and illuminated growth and processing strategies, including how to cool from the processing temperature, depend entirely on the specifics of the semiconductor and its ensemble of native defects and any extrinsic alloying or doping species. The system we have considered herein is exceedingly simplified in order to interrogate the general response of a single hypothetical defect population to the presence of excess carriers. Results in real materials for a particular defect may be more muted or completely different because of interactions of the whole ensemble of defect types with the excess carriers. For example, the response of one defect charge state might be constrained by Fermi level pinning induced by another defect type, thus adding complications to the general rules laid out herein. Furthermore, while the addition of a new chemical potential to vary during processing opens the possibility of solving beguiling material challenges, it also adds complications. For example, a corollary can be drawn in a case where both compensating and same-polarity native defects are of concern. Processing with excess carriers at some elevated temperature might be used to lock-in a lower concentration of compensating defects, but slow cooling in the light may enhance the same-polarity defect. In that case more complex processing strategies might be required: rapid quenching through lower temperatures under dark conditions might be required in order to suppress the same-polarity defects. These are just some first examples of how understanding the effects of excess carriers can complicate processing strategies but yield more control over the end properties of semiconductors.

**Summary and Conclusions**

In summary, we have developed a general model that accounts for changes to the defect formation energy in the presence of non-equilibrium carrier concentrations. The Fermi level component of the formation enthalpy of a defect with charge $q$ is represented as an admixture of the electron and hole quasi-Fermi levels (QFLs) based on the relative electron and hole emission and capture rates of defects with charges $q$-1 and $q$+1. The resulting effects were studied theoretically in a system defined by a ZnSe-like wide-bandgap semiconductor, one possible donor-like native defect having $X^0$ and $X^{+1}$ charge states, and different concentrations of p- or n-type extrinsic dopants. We find that that non-equilibrium carrier concentrations have the potential to reduce the concentration of compensating defects but can in some circumstances cause an increased concentration of defects with the same charge sign as the extrinsic dopant. The later scenario is a result of low electron or hole emission rates, which alter the dependence of formation enthalpy on the electron or hole QFL.  Increases in the defect concentration are expected to be pronounced in wide bandgap semiconductors, where the charge transition levels can be quite far from the relevant band edge.  This theory predicts how non-equilibrium carrier concentrations can be used to manipulate defect populations during semiconductor growth and processing. Injection of excess carriers via illumination, radiation, or current injection could especially be used to lower the concentration of compensating defects and overcome doping limits that otherwise restrict the technological use of certain semiconductors.  Also, this mechanism may have implications on the evolution of defect populations over time in devices.  In LEDs, lasers, solar cells, or power devices the coupling of the non-equilibrium carriers vital to their operation to the charged defect populations may cause changes over time considered as degradation, "burn-in", or even enhancement of performance.  These changes would not be capable of being understood within a framework of equilibrium and thus this model may also explain some heretofore unexplained effects of operating conditions.

  Lastly, we conclude with suggestions for definitely experimentally testing the hypotheses of this particular theoretical mechanism, a task that may be extremely challenging due to a number of confounding alternative physical mechanisms.  Not all experiments involving illumination during epitaxial growth or annealing should be considered to have tested this particular mechanism, even if the experiments result in desirable empirical outcomes consistent with the theory's predictions.  Many other physical effects can result from illumination during semiconductor epitaxial growth, including adatom desorption, modification of atomic surface diffusion and attachment at steps, electric field screening, improved material quality, and changes in the incorporation of certain alloy species.[16,17,18,19,20,21,22] Particularly in the case of chemical vapor deposition (i.e. OMVPE / MOCVD), potential confounding photochemical reactions may occur that are unrelated to the mechanism detailed herein.[23,24]  Given that kinetics at the growth interface govern most epitaxial growth processes, we suggest that annealing experiments in the presence of well-controlled overpressures of relevant atomic or molecular species which do not absorb the wavelength used will provide the best measure of the effect of non-equilibrium carrier concentrations on defect populations. Molecular beam epitaxy (MBE) chambers may provide such facilities, as well as the ability to assess whether or not desorption or growth is occurring using RHEED and/or quantitative mass spectrometry (although high energy electrons also will produce excess carrier populations).[25] Semiconductors with well-understood native and doping-related defect ensembles would be ideal starting candidates. The effects of the defect charge transition energy on the overall change in defect population should also be tested by examining changes in both narrow and wide bandgap semiconductors. Such experimental investigations are critical for distinguishing this mechanism from other kinetically-controlled defect formation pathways.

## Acknowledgements

K.A. and M.A.S. contributed equally to the development of the model, the generation and analysis of the results and the composition and revision of the manuscript. K.A.

acknowledges financial support from the Department of Energy Office of Science, Basic Energy Sciences under contract DE-AC36-08GO28308. M.A.S. acknowledges support from the Department of Energy through the Bay Area Photovoltaic Consortium award DE-EE0004946.**References**

[1] Walukiewicz, W., Mechanisms of Fermi-level stabilization in semiconductors, *Phys. Rev. B*, **37**, 4760-4763 (1988).

[2] Zhang, S.B., Wei, S.-H. and Zunger, A. Microscopic origin of the phenomenological equilibrium "doping limit rule" in n-type III-V semiconductors, *Phys. Rev. Lett.*, **84**, 1232-1235 (2000).

[3] Qiu, J., DePuydt, J.M., Cheng, H. and Haase, M.A. Heavily doped p-ZnSe:N grown by molecular beam epitaxy, *Appl. Phys. Lett.*, **59**, 2992 (1991).

[4] Ohki, A., Kawaguchi, Y., Ando, K. and Katui, A. Acceptor compensation mechanism by midgap defects in nitrogen-doped ZnSe films, *Appl. Phys. Lett.*, **59**, 671 (1991).

[5] Van de Walle, C.G. and Neugebauer, J. First-principles calculations for defects and impurities: Applications to III-nitrides, *J. Appl. Phys.*, **95**, 3851 (2004).

[6] Polyakov, A.Y. and Lee, I.-H. Deep traps in GaN-based structures as affecting the performance of GaN devices, *Materials Science and Engineering R*, **94**, 1 (2015)

[7] Ozgur, U., Alivov, Ya.I., Liu, C., Teke, A., Reshchikov, M.A., Dogan, S., Avrutin, V., Cho, S.-J. and Morkoc, H. A comprehensive review of ZnO materials and devices, *J. Appl. Phys.*, **98**, 041301 (2005).

[8] K. Alberi and M.A. Scarpulla, Suppression of compensating native defect formation during semiconductor processing via excess carriers, *Scientific Reports*, **6**, 27954 (2016)

[9] Zhang, S.B., and Northrup, J.E., Chemical potential dependence of defect formation energies in GaAs: application to Ga self-diffusion, *Phys. Rev. Lett.*, **67**, 2339 (1991)

[10] Shockley, W. & Read, W.T. Statistics of the recombination of holes and electrons. *Phys. Rev.* **87**, 835-842 (1952).

[11] Hall, R.N. Electron-hole recombination in Germanium. *Phys. Rev.* **87**, 387 (1952)23

**Supplementary Material: Effects of excess carriers on native defects in wide bandgap semiconductors**

Kirstin Alberi[1,*] and Michael A. Scarpulla[2,3]
[1] National Renewable Energy Laboratory, Golden, CO, USA
[2] Materials Science & Engineering, University of Utah, Salt Lake City, UT, USA
[3] Electrical & Computer Engineering, University of Utah, Salt Lake City, UT, USA

This Supplementary Material Section includes the full results of our defect calculations over a wider range of doping concentrations as well as explicit free electron and hole concentrations for reference.

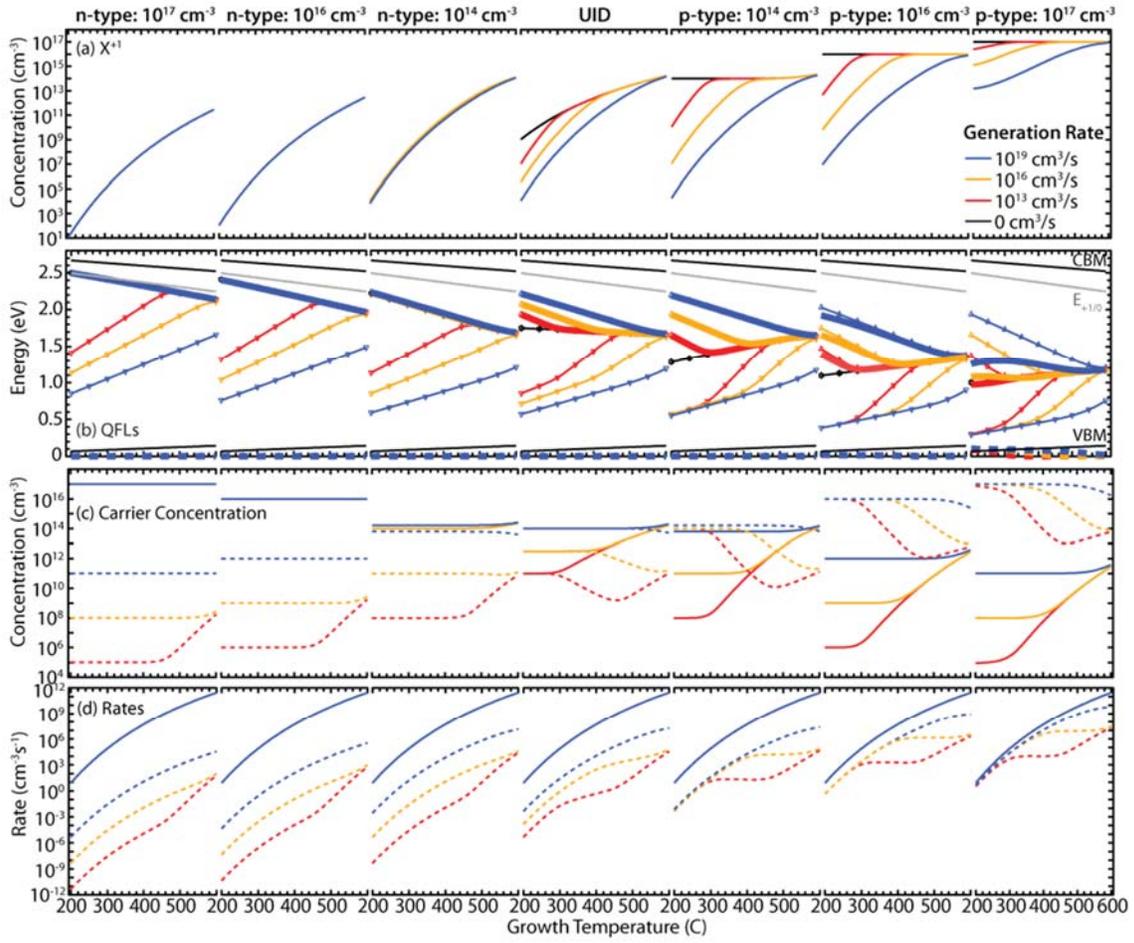
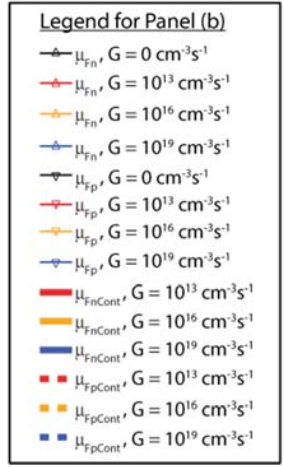
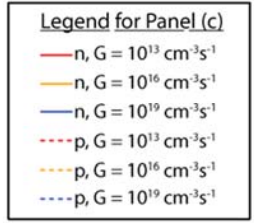
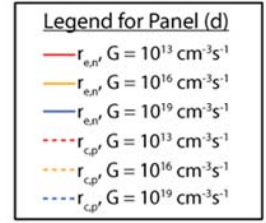

**Fig. S1** (Color online) Calculated parameters for Case 1 in which $E_{+1/0}$ is positioned close to $E_c$. Calculations were carried out as a function of temperature, extrinsic doping concentration (noted at the top of the figure) and photogeneration rate, $G$. UID stands for unintentionally doped. (a) Concentration of $X^{+1}$ defects, (b) $\mu_{Fn}$ and $\mu_{Fp}$ along with the contribution of $\mu_{Fn}$ (denoted $\mu_{FnCont}$) and $\mu_{Fp}$ (denoted $\mu_{FpCont}$) to $\mu_F$ used in the defect formation energy calculation. The CBM, VBM and $E_{+1/0}$ values are also marked. (c) Electron and hole carrier concentrations. (d) Electron emission and hole capture rates. The legends for panels b-d are shown at the bottom of the figure.

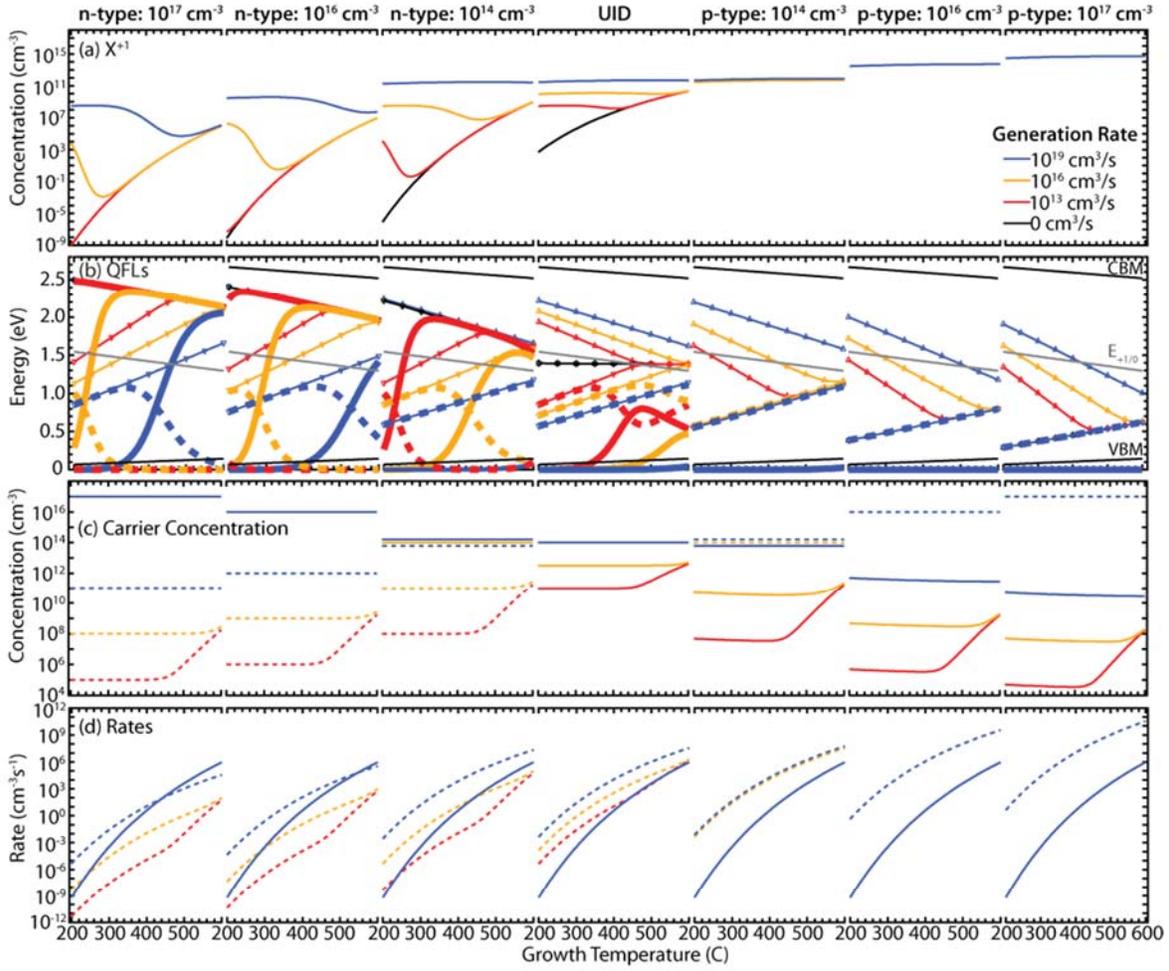
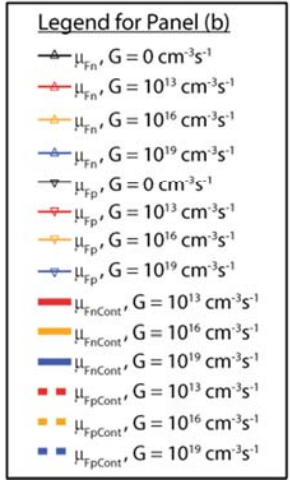
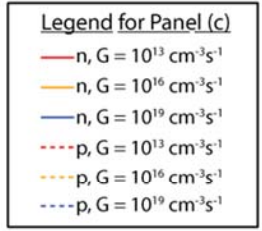
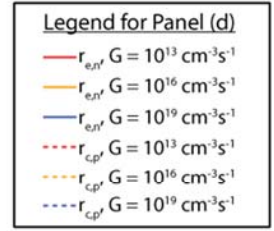

**Fig. S2** (Color online) Calculated parameters for Case 1 in which $E_{+1/0}$ is positioned close to $E_c$. Calculations were carried out as a function of temperature, extrinsic doping concentration (noted at the top of the figure) and photogeneration rate, $G$. UID stands for unintentionally doped. (a) Concentration of $X^{+1}$ defects, (b) $\mu_{Fn}$ and $\mu_{Fp}$ along with the contribution of $\mu_{Fn}$ (denoted $\mu_{FnCont}$) and $\mu_{Fp}$ (denoted $\mu_{FpCont}$) to $\mu_F$ used in the defect formation energy calculation. The CBM, VBM and $E_{+1/0}$ values are also marked. (c) Electron and hole carrier concentrations. (d) Electron emission and hole capture rates. The legends for panels b-d are shown at the bottom of the figure.

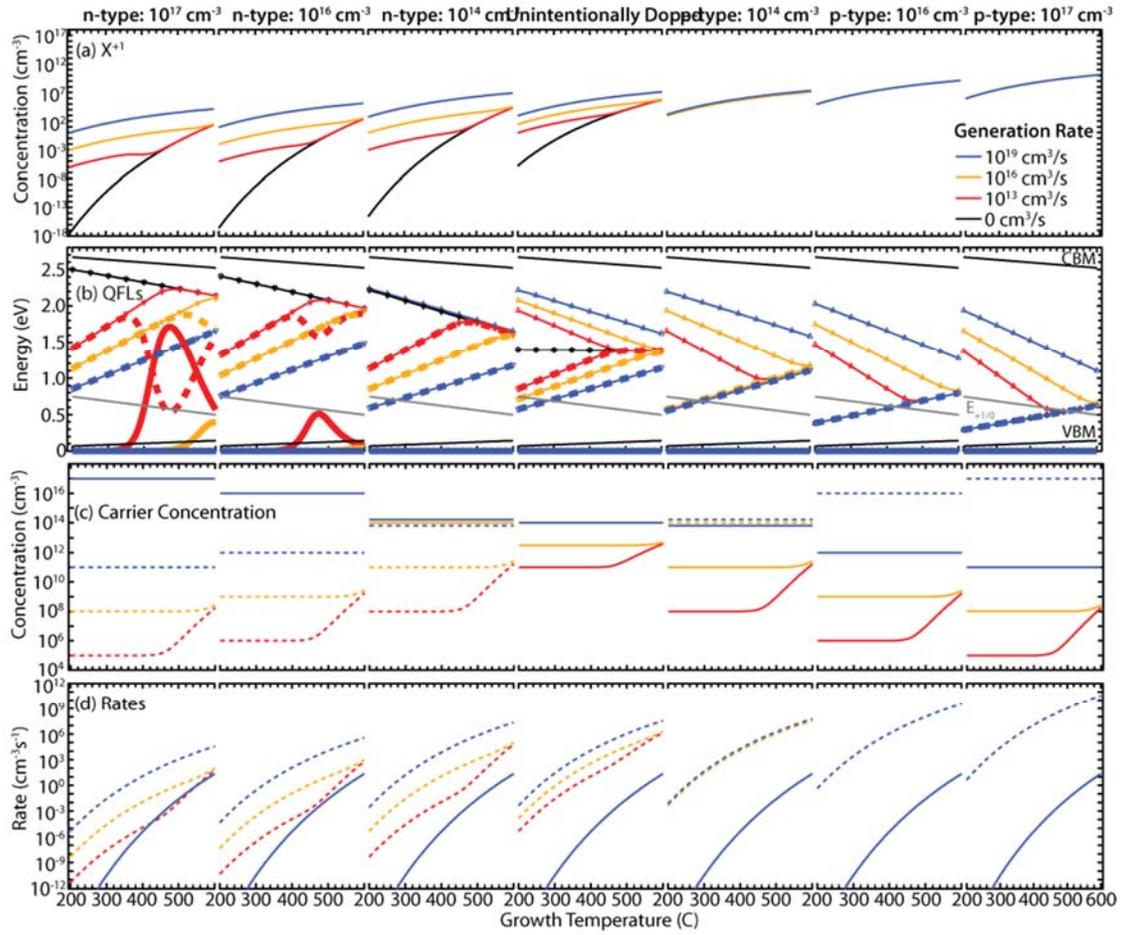
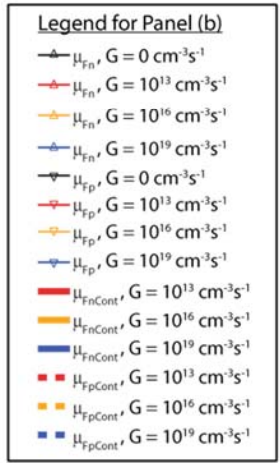
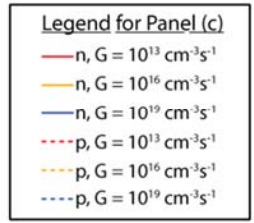
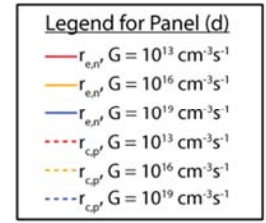

**Fig. S3** (Color online) Calculated parameters for Case 1 in which $E_{+1/0}$ is positioned close to $E_c$. Calculations were carried out as a function of temperature, extrinsic doping concentration (noted at the top of the figure) and photogeneration rate, $G$. UID stands for unintentionally doped. (a) Concentration of $X^{+1}$ defects, (b) $\mu_{Fn}$ and $\mu_{Fp}$ along with the contribution of $\mu_{Fn}$ (denoted $\mu_{FnCont}$) and $\mu_{Fp}$ (denoted $\mu_{FpCont}$) to $\mu_F$ used in the defect formation energy calculation. The CBM, VBM and $E_{+1/0}$ values are also marked. (c) Electron and hole carrier concentrations. (d) Electron emission and hole capture rates. The legends for panels b-d are shown at the bottom of the figure.